\pdfoutput=1
\documentclass[sigconf]{acmart}
\copyrightyear{2018}
\acmYear{2018}
\setcopyright{acmcopyright}
\acmConference[WSDM 2018]{Eleventh ACM International
Conference on Web Search and Data Mining}{}{February 5--9, 2018, Marina Del Rey, CA, USA}
\acmPrice{15.00}
\acmDOI{10.1145/3159652.3159656}
\acmISBN{978-1-4503-5581-0/18/02}

\usepackage{booktabs} 

\usepackage{subcaption,verbatim,engord,enumitem,multirow,bm,caption}
\usepackage{algorithm,algorithmic}
\usepackage{amsmath,dsfont}
\usepackage{amssymb}
\usepackage{graphicx}
\usepackage{amscd}
\usepackage{balance}

\begin{document}
\title{Personalized Top-N Sequential Recommendation
via Convolutional Sequence Embedding}

\author{Jiaxi Tang}
\affiliation{%
  \institution{School of Computing Science}
  \streetaddress{Simon Fraser University}
  \city{British Columbia} 
  \state{Canada} 
}
\email{jiaxit@sfu.ca}

\author{Ke Wang}
\affiliation{%
  \institution{School of Computing Science}
  \streetaddress{Simon Fraser University}
  \city{British Columbia} 
  \state{Canada} 
}
\email{wangk@cs.sfu.ca}

\fancyhead{} 

\begin{abstract}
Top-$N$ sequential recommendation models
each user as a sequence of items interacted in the past 
and aims to predict top-$N$ ranked items that a user
will likely interact in a ``near future''. The order of interaction implies that sequential patterns play an important role where more recent items in a sequence have a larger impact on the next item. In this paper, we propose a Convolutional Sequence Embedding Recommendation Model (\emph{Caser}) as a solution to address this requirement. 
The idea is to  
embed a sequence of recent items into an ``image'' in the time and latent spaces
and learn sequential patterns as local features of the image using convolutional filters. This approach provides a unified and flexible network structure for capturing both general preferences and sequential patterns. The experiments on public data sets demonstrated that Caser consistently outperforms state-of-the-art sequential recommendation methods on a variety of common evaluation metrics.
\end{abstract}

%
%

\ccsdesc[500]{Information systems~Retrieval models and ranking}


\keywords{Recommender System; Sequential Prediction; Convolutional Neural Networks }

\maketitle

\section{introduction}\label{sec:intro}
Recommender systems have become a core technology in many applications.
Most systems, e.g., \emph{top-$N$ recommendation}  \cite{hu2008collaborative}\cite{pan2008one},
recommend the items based on the user's \emph{general preferences}
without paying attention to the recency of items.

For example, some user always prefer Apple's products to Samsung's products. General preferences represent user's long term and static behaviors. Another type of user behaviors is \emph{sequential patterns} where the next item or action more likely depends on the items or actions the user engaged recently. Sequential patterns represent the user's short term and dynamic behaviors
and come from a certain relationship between the items within a close proximity of time. For example, a user likely buys phone accessories soon after buying an iPhone, though in general the user does not buy phone accessories. In this case, the systems that consider only general preferences will miss the opportunity of recommending phone accessories after selling an iPhone since buying phone accessories is not a long term user behavior.

\subsection{Top-$N$ Sequential Recommendation}
To model user's sequential patterns, the work in \cite{rendle2010factorizing, liu2009hybrid} considers \emph{top-$N$ sequential recommendation} that recommends $N$ items that a user likely interacts with in a near future. This problem assumes a set of users $\mathcal{U} = \{u_1, u_2,\cdots,u_{|\mathcal{U}|}\}$ and a universe of items $\mathcal{I} = \{i_1, i_2,\cdots,i_{|\mathcal{I}|}\}$. Each user $u$ is associated with a sequence of some items from $\mathcal{I}$, $\mathcal{S}^u=(\mathcal{S}^u_1, \cdots, \mathcal{S}^u_{|\mathcal{S}^u|})$, where $\mathcal{S}^u_i \in \mathcal{I}$.
The index $t$ for $\mathcal{S}^u_t$ denotes the order in which an action occurs in the sequence $\mathcal{S}^u$, not the absolute timestamp as in temporal recommendation like~\cite{wu2017rnn,zhang2014latent,koren2010collaborative}.
Given all users' sequences $\mathcal{S}^u$, the goal is to recommend each user a list of items that maximize her/his future needs, by considering both general preferences and sequential patterns. Unlike conventional top-$N$ recommendation,
top-$N$ sequential recommendation models the user behavior as a sequence of items, instead of a set of items.


\begin{figure*}[t!]  

        \centering
        \begin{subfigure}[b]{0.25\textwidth}
                \includegraphics[width=\textwidth]{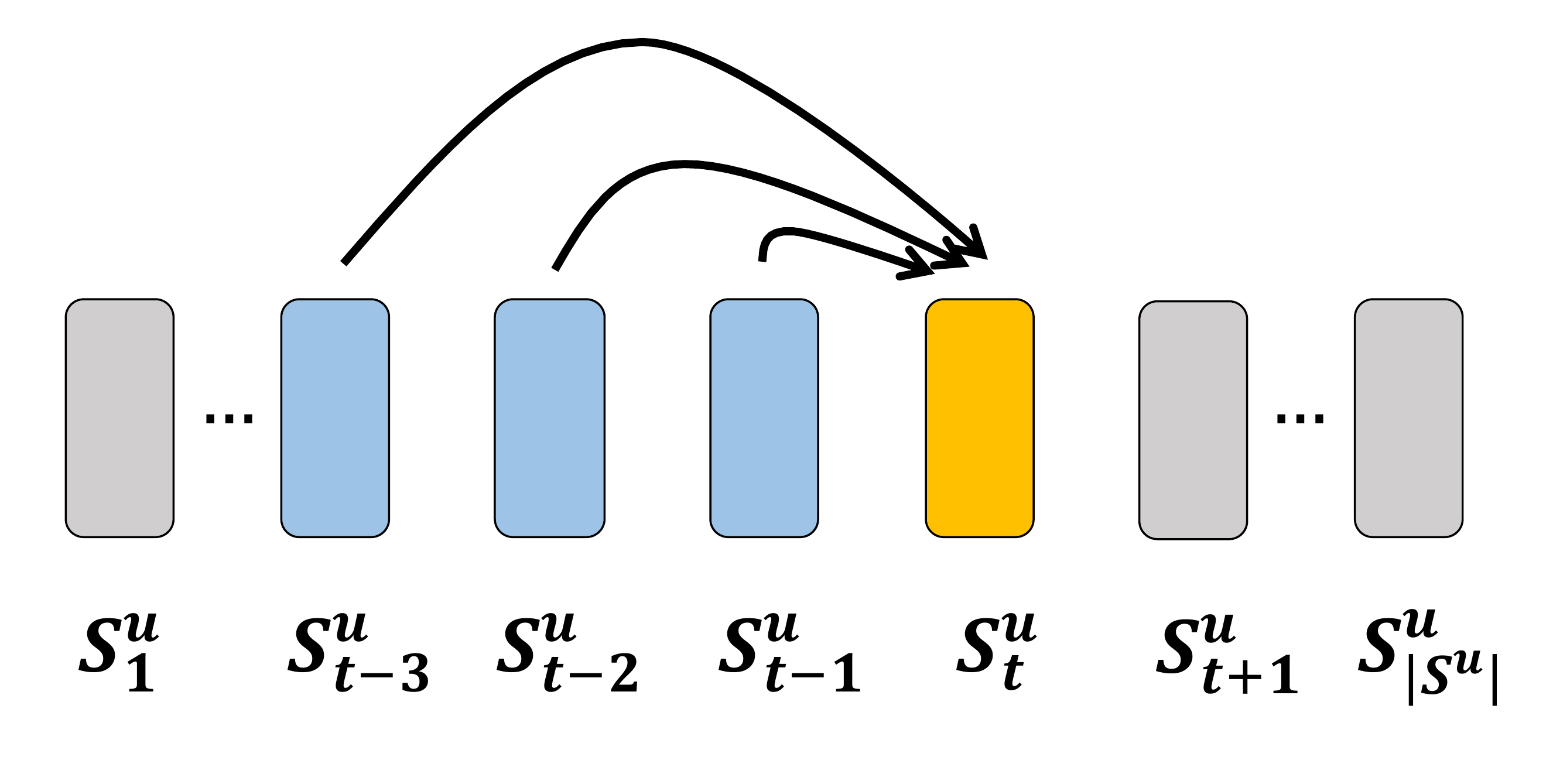}
                \vspace{-0.5cm}
                \caption{point-level}
                \label{fig:demo_pl}
        \end{subfigure}%
        ~
        \hspace{0.02in}
        \begin{subfigure}[b]{0.25\textwidth}
                \includegraphics[width=\textwidth]{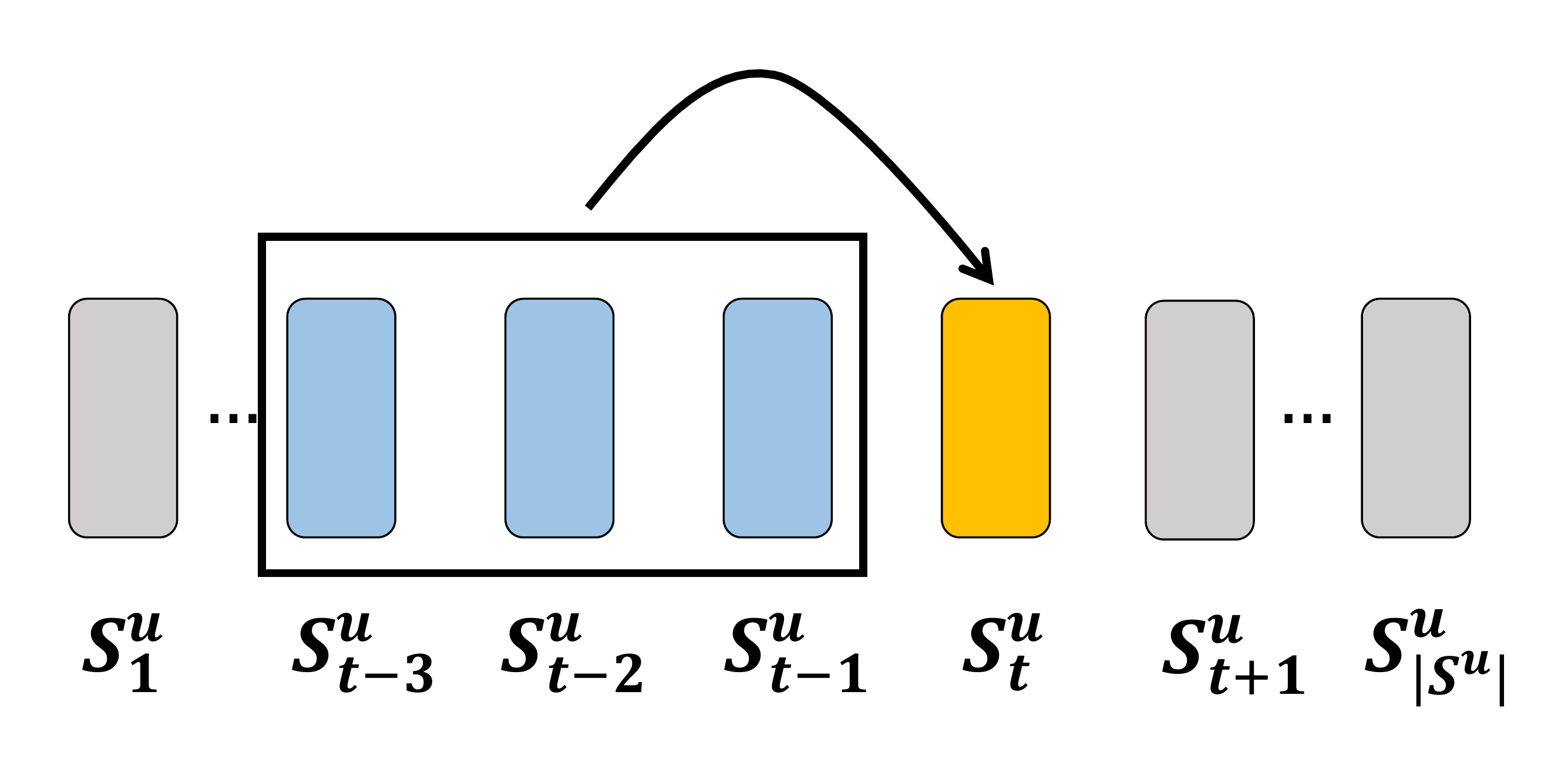}
                \vspace{-0.5cm}
                \caption{union-level, no skip}
                \label{fig:demo_ul1}
        \end{subfigure}
        ~
        \hspace{0.02in}
        \begin{subfigure}[b]{0.25\textwidth}
                \includegraphics[width=\textwidth]{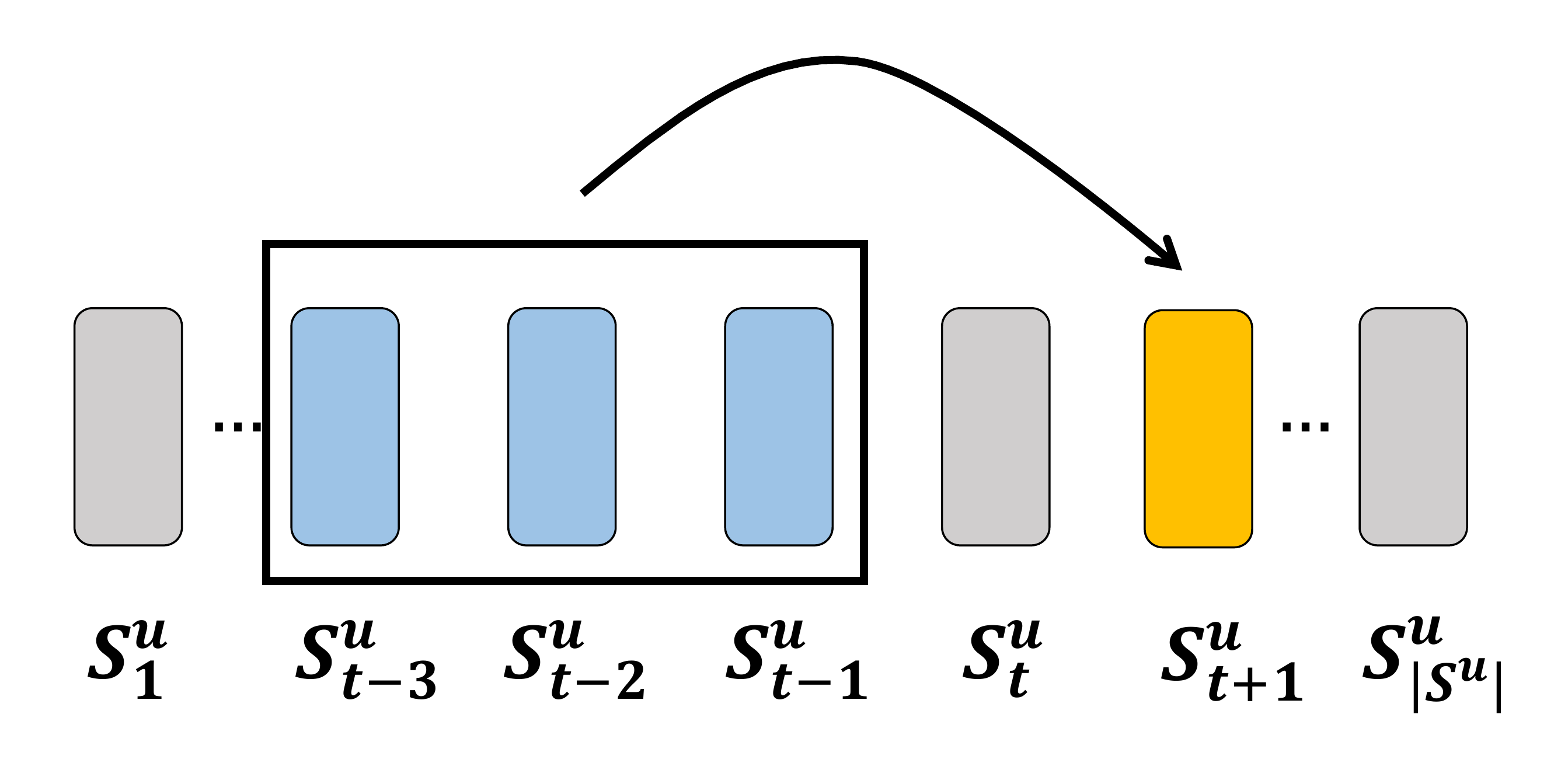}
                \vspace{-0.5cm}
                \caption{union-level, skip once}
                \label{fig:demo_ul2}
        \end{subfigure}
        \vspace{-0.4cm}
        \caption{An example of point and union level dynamic pattern influences, the order of Markov chain $L=3$}\label{fig:demo}

\end{figure*}


\vspace{-0.2cm}
\subsection{Limitations of Previous Work}
The Markov chain based model~\cite{rendle2010factorizing, HeMcA16b, cheng2013you,wang2015learning}
is an early approach to top-$N$ sequential recommendation, where an $L$-order Markov chain makes recommendations based on $L$ previous actions.
The first-order Markov chain is an item-to-item transition matrix learnt using maximum likelihood estimation. Factorized personalized Markov chains~(FPMC)~\cite{rendle2010factorizing} proposed by Rendle \emph{et al.} and its variant~\cite{cheng2013you} improved this method by factorizing this transition matrix into two latent and low-rank sub-matrices.
Factorized Sequential Prediction with Item
Similarity ModeLs~(Fossil)~\cite{HeMcA16b} proposed by He \emph{et al.} generalizes this method to high-order Markov chains using a weighted sum aggregation over previous items' latent representations.
However, existing approaches suffered from two major limitations:

\noindent{\textbf{Fail to model union-Level sequential patterns}}. As shown in Figure~\ref{fig:demo_pl}, the Markov chain models only \textbf{point-level} sequential patterns where each of the previous actions (blue) influences the target action (yellow) individually, instead of collectively.
FPMC and Fossil fall into this taxonomy. Although
Fossil~\cite{HeMcA16b} considers a high-order Markov chain, the overall influence is a weighted sum of previous items' latent representations factorized from first-order Markov transition matrices. Such aggregation of point-level influences is not sufficient to model the \textbf{union-level} influences shown in Figure~\ref{fig:demo_ul1} where several previous actions, in that order, jointly influence the target action. For example, buying both milk and butter together leads to a higher probability of buying flour than buying milk or butter individually; buying both RAM and Hard Drive is a better indication of buying Operating System next than buying only one of the components.


\noindent{\textbf{Fail to allow skip behaviors}}. Existing models don't consider \textbf{skip behaviors} of sequential patterns as shown in Figure~\ref{fig:demo_ul2}, where the impact
from past behaviors may skip a few steps and still have strength.
For example, a tourist has check-ins sequentially at airport, hotel, restaurant, bar, and attraction. While the check-ins at the airport and hotel do not
immediately precede the check-in of the attraction, they are strongly associated with the latter. On the other hand, the check-in at the restaurant or bar has little influence on the check-in of the attraction (because they do not necessarily occur).
A $L$-order Markov chain does not explicitly model
such skip behaviors because it assumes that the $L$ previous
steps have an influence on the immediate next step.

\begin{figure}[h!]  

        \centering
        \begin{subfigure}[b]{0.265\textwidth}
                \includegraphics[width=\textwidth]{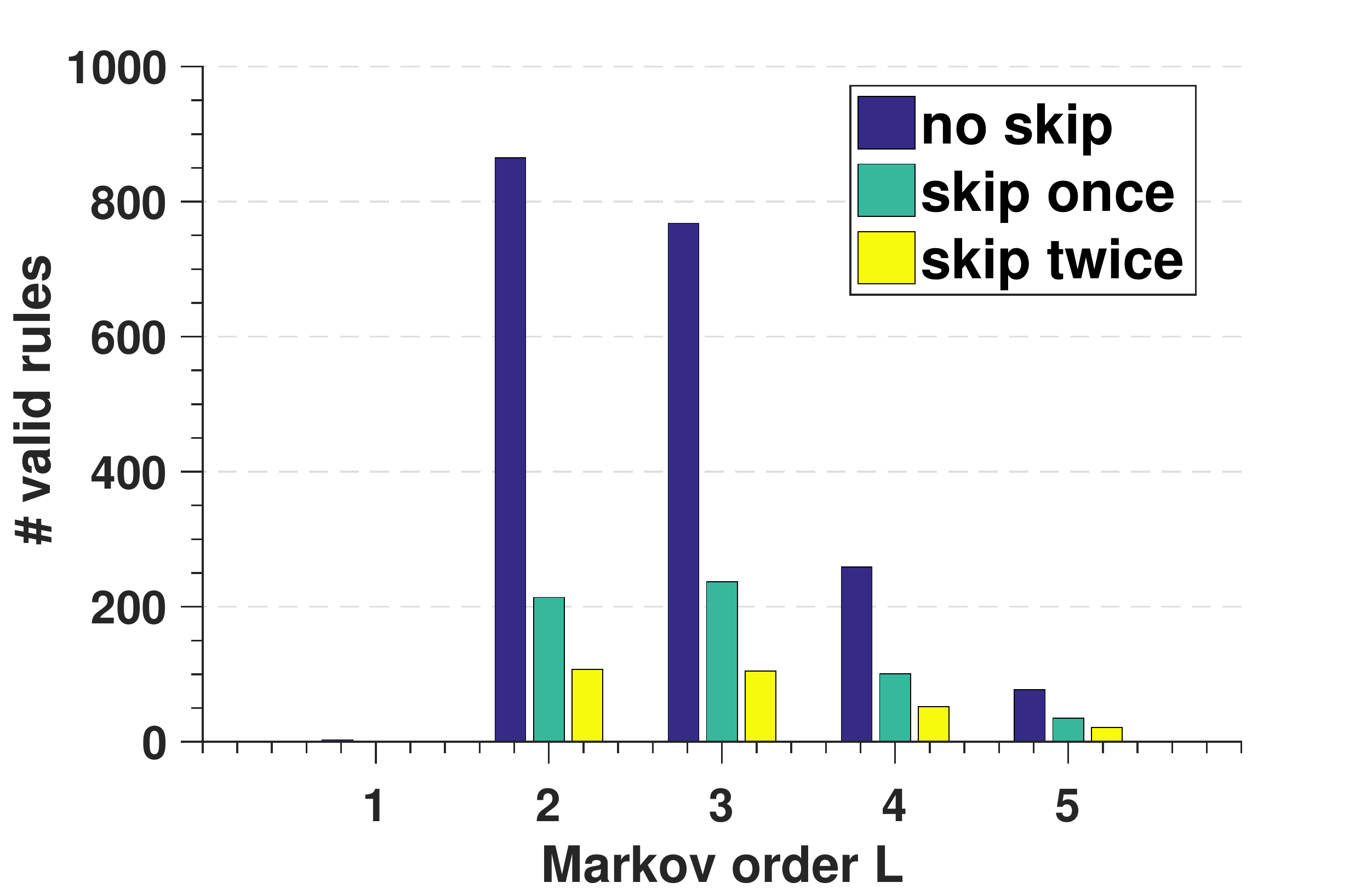}
                \vspace{-0.4cm}
                \caption{MovieLens}
                \label{fig:obv_ml}
        \end{subfigure}%
        ~
        \hspace{-0.247in}
        \begin{subfigure}[b]{0.265\textwidth}
                \includegraphics[width=\textwidth]{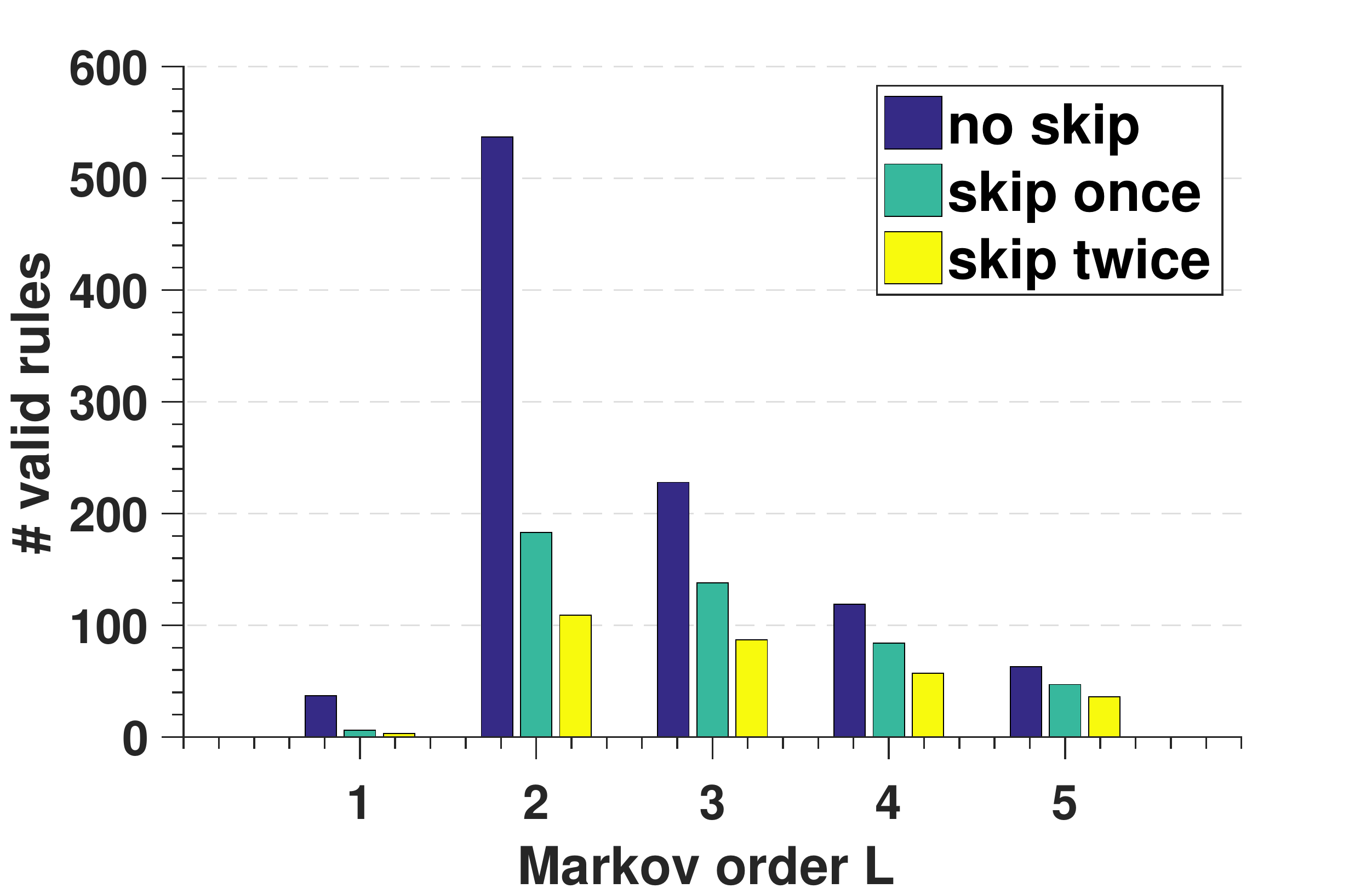}
                \vspace{-0.4cm}
                \caption{Gowalla}
                \label{fig:obv_gw}
        \end{subfigure}
        \vspace{-0.7cm}
        \caption{The number of association rules vs $L$ and skip steps. The minimum support count = 5 and the minimum confidence = 50\%.}\label{fig:obv}
\end{figure}

To provide evidences of union-level influences and skip behaviors, 
we mine sequential association rules~\cite{agrawal1995mining, han2011data} of the following form from two real life data sets, MovieLens and Gowalla (see the details of these data sets in Section~\ref{sec:exper})
\begin{equation}\label{equ:rules}
(\mathcal{S}^u_{t-L},\cdots,\mathcal{S}^u_{t-2},\mathcal{S}^u_{t-1}) \rightarrow \mathcal{S}^u_{t}.
\end{equation}
For a rule $X\rightarrow Y$ of the above form, the support count $sup(XY)$ is the number of sequences in which $X$ and $Y$ occur in order as in the rule, and the confidence, $\frac{sup(XY)}{sup(X)}$, is the percentage of the sequences in which $Y$ follows $X$ among those in which $X$ occurs.
This rule represents the joint influence of all the items in $X$ on $Y$. By changing the right hand side to $\mathcal{S}^u_{t+1}$ or $\mathcal{S}^u_{t+2}$, the rule also captures the influences with one or two step skips.
Figure~\ref{fig:obv} summarizes the number of rules found versus the Markov order $L$ and skip steps with the minimum support count = 5 and the minimum confidence = 50\% (we also tried the minimum confidence of 10\%, 20\%, and 30\%, these trends are similar). Most rules have the orders $L=2$ and $L=3$ and the confidence of rules gets higher for larger $L$. The figure also tells that a sizable number of rules have skip steps 1 or 2.
These findings support the existence of union-level influences and skip behaviors.

\subsection{Contributions}
To address these above limitations of existing works, we propose a \emph{ConvolutionAl Sequence Embedding Recommendation Model}, or \emph{Caser} for short, as a solution to top-$N$ sequential recommendation. This model leverages the recent success of convolution filters of Convolutional Neural Network (CNN) to capture local features for image recognition~\cite{krizhevsky2012imagenet,karpathy2014large} and natural language processing~\cite{yoon14convolution}. The novelty of Caser is to represent the previous $L$ items as an $L\times d$ matrix $\boldsymbol{E}$, where $d$ is the number of latent dimensions and the rows preserve the order of the items. Similar to~\cite{yoon14convolution}, we regard this embedding matrix as the ``image'' of the $L$ items in the latent space and search for sequential patterns as local features of this ``image'' using various convolutional filters. Unlike image recognition, however, this ``image'' is not given in the input and must be learnt simultaneously with all filters.

\begin{figure*}[!t]  
\centering
\includegraphics[scale=0.15]{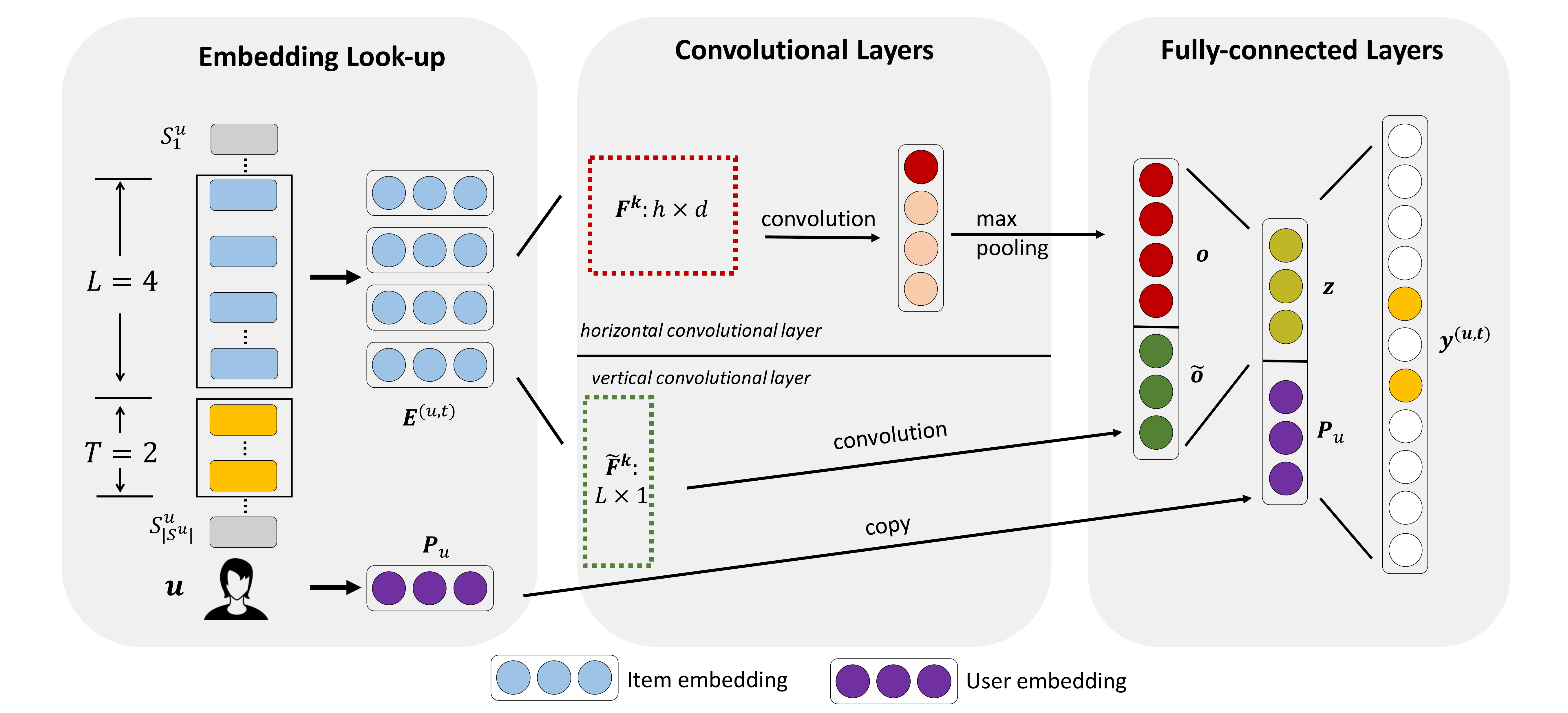}
\vspace{-0.4cm}
\caption{The network architecture of \emph{Caser}. The rectangular boxes represent items $\mathcal{S}^u_1,\cdots,\mathcal{S}^u_{|\mathcal{S}^u|}$ in user sequence, whereas a rectangular box with circles inside stands for a certain vector \emph{e.g.,} user embedding $\boldsymbol{P}_u$. The dash rectangular boxes are convolutional filters with different sizes. The red circles in convolutional layers stand for the max values in each of the convolution results. Here we are using previous 4 actions ($L=4$) to predict which items this user will interact with in next 2 steps ($T=2$).}
\label{fig:network}
\end{figure*}

Compared to existing methods, Caser offers several distinct advantages. (1) Caser uses horizontal and vertical convolutional filters to capture sequential patterns at point-level, union-level, and of skip behaviors. (2)
Caser models both users' general preferences and sequential patterns, and generalizes several existing state-of-the-art methods in a single unified framework.
(3) Caser outperforms state-of-the-art methods for top-$N$ sequential recommendation on real life data sets.
In the rest of the paper, we discuss further related work in Section~\ref{sec:related}, the Caser method in Section~\ref{sec:model}, and experimental studies in Section~\ref{sec:exper}.

\section{Further Related Work}\label{sec:related}

Conventional recommendation methods, \emph{e.g.}, collaborative filtering~\cite{sarwar2001item}, matrix factorization~\cite{koren2009matrix,salakhutdinov2007probabilistic}, and top-$N$ recommendation \cite{hu2008collaborative}\cite{pan2008one}, are not suitable for capturing sequential patterns because they do not model the order of actions.
Early works on sequential pattern mining~\cite{agrawal1995mining, han2011data} find explicit sequential association rules based on statistical co-occurrences
~\cite{liu2009hybrid}. This approach depends on the explicit representation of patterns, thus, could miss patterns in unobserved states. Also, it suffers from a potentially large search space, sensitivity to threshold settings, and a large number of rules, most being redundant.



Restricted Bolzmann Machine (RBM)~\cite{salakhutdinov2007restricted} is the first successful 2-layers neural network that is applied to recommendation problems. Auto-encoder framework~\cite{sedhain2015autorec, wang2015collaborative} and its variant denoising auto-encoder~\cite{wu2016collaborative} also produce a good recommendation performance. Convolutional neural network (CNN)~\cite{zheng2017jdm} has been used to extract users' preferences from their reviews. None of these works is for sequential recommendation.



Recurrent neural networks~(RNN) was used for session-based recommendation \cite{hidasi2015session,jannach2017recurrent}. While RNN has shown to have an impressive capability in modeling sequences~\cite{mikolov2010recurrent}, its sequentially connected network structure may not work well under sequential recommendation setting. Because in sequential recommendation problem, not all adjacent actions have dependency relationships (\emph{e.g.} a user bought $i_2$ after $i_1$ only because she loves $i_2$). Our experimental results in Section~\ref{sec:exper} verify this point: RNN-based method performs better when data sets contains considerable sequential patterns. 
While our proposed method doesn't model sequential pattern as adjacent actions, it adopts convolutional filters from CNN and model sequential patterns as local features of the embeddings of previous items. This approach offers the flexibility of modeling sequential patterns at both point level and union level, and skip behaviors in a single unified framework. In fact, we will show that Caser generalizes several state-of-the-art methods.

A related but different problem is temporal recommendation \cite{zhang2014latent, wu2017rnn,song2016multi}.
For example, temporal recommendation recommends coffee in the morning, instead of evening, whereas our top-$N$ sequential recommendation would recommend phone accessories soon after a user bought an iPhone, independently of the time. Clearly, the two problems are different and require different solutions.

\section{Proposed Methodology}\label{sec:model}
The proposed model, ConvolutionAl Sequence Embedding Recommendation (\emph{Caser}),
incorporates the Convolutional Neural Network (CNN) to learn sequential
features, and Latent Factor Model~(LFM) to learn user specific features.
The goal of Caser's network design is multi-fold: capture both user's general preferences and sequential patterns, at both union-level and point-level, and capture skip behaviors, all in unobserved spaces.
Shown in Figure~\ref{fig:network} Caser
consists of three components: Embedding Look-up, Convolutional Layers, and Fully-connected Layers. To train the CNN, for each user $u$,
we extract every $L$ successive items as input and their next $T$ items as the targets from the user's sequence $\mathcal{S}^u$, shown on the left side of Figure~\ref{fig:network}.
This is done by sliding a window of size $L+T$ over the user's sequence, and each window generates a training instance for $u$, denoted by a triplet ($u$, previous $L$ items, next $T$ items).


\subsection{Embedding Look-up} %
\emph{Caser} captures sequence features in the latent space by feeding the embeddings of previous $L$ items into the neural network. The embedding $\boldsymbol{Q}_i \in \mathbb{R}^d$ for item $i$ is a similar concept to its latent factors. Here $d$ is the number of latent dimensions. The embedding look-up operation retrieves the previous $L$ items' embeddings 
and stacks them together, resulting in a matrix $\boldsymbol{E}^{(u,t)} \in \mathbb{R}^{L \times d}$ for user $u$ at time step $t$:
\begin{equation}\label{equ:embed}
\begin{aligned}
\boldsymbol{E}^{(u,t)} = \left[ \begin{matrix} \boldsymbol{Q}_{\mathcal{S}^u_{t-L}} \\\vdots\\  \boldsymbol{Q}_{\mathcal{S}^u_{t-2}}\\ \boldsymbol{Q}_{\mathcal{S}^u_{t-1}}\end{matrix} \right].
\end{aligned}
\end{equation}
Along with the item embeddings, we also have an embedding $\boldsymbol{P}_u \in \mathbb{R}^d$ for a user $u$, representing user features in latent space. These embeddings are represented by blue and purple circles in the box of Embedding Look-up in Figure~\ref{fig:network}.


\subsection{Convolutional Layers}\label{sec:conv} 

\begin{figure}[t]  
\centering
\includegraphics[scale=0.109]{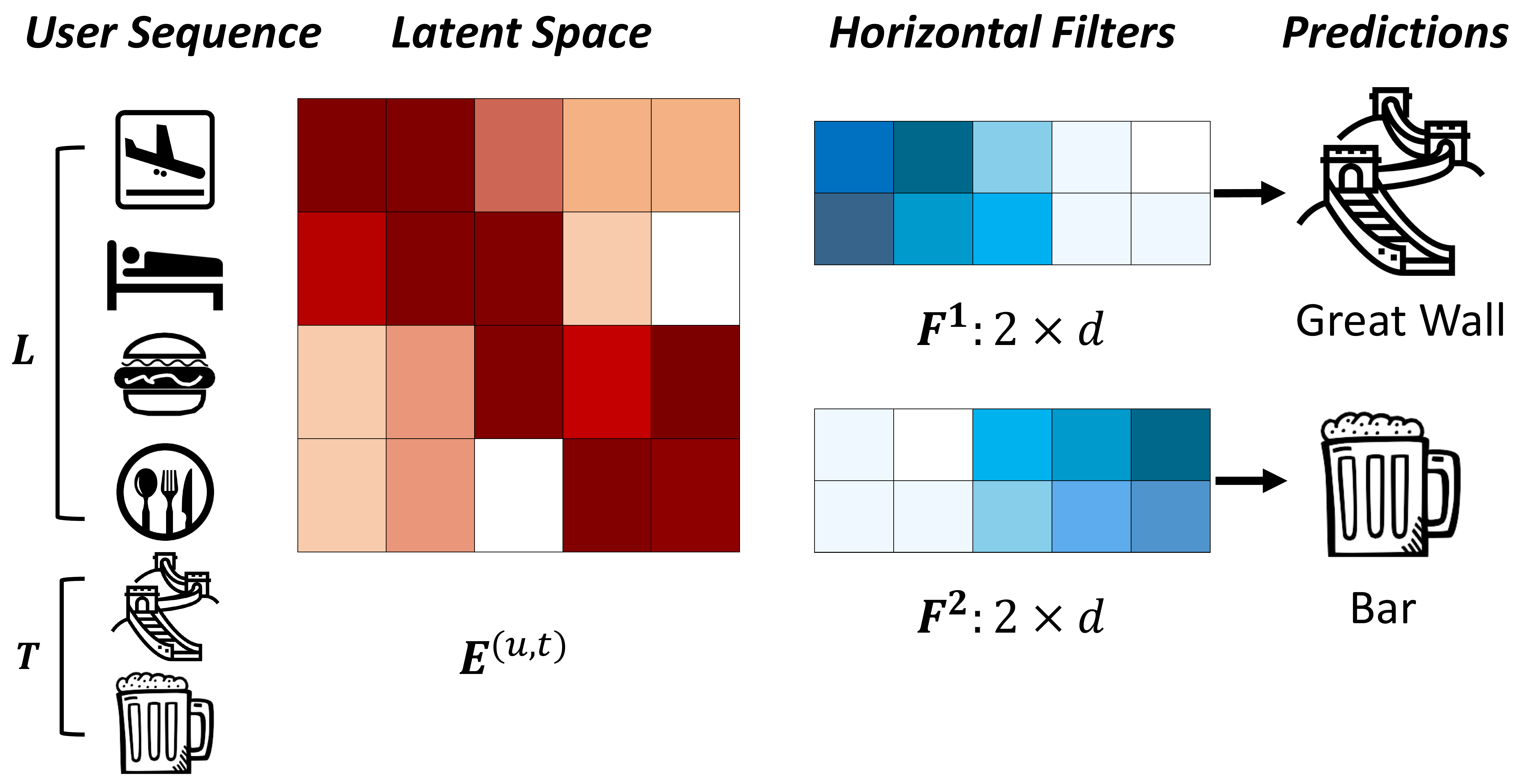}
\vspace{-0.4cm}
\caption{Darker colors mean larger values. The first filter captures ``(Airport, Hotel) $\rightarrow$ Great Wall'' by interacting with the embedding of airport and hotel and skipping that of fast food and restaurant. The second filter captures ``(Fast Food, Restaurant) $\rightarrow$ Bar''.}
\label{fig:hcase}
\end{figure}

Our approach leverages the recent success of convolution filters of CNN in capturing local features for image  recognition~\cite{krizhevsky2012imagenet,karpathy2014large} and natural language processing~\cite{yoon14convolution}. Borrows the idea of using CNN in text classification~\cite{yoon14convolution}, our approach regards the $L\times d$ matrix $\boldsymbol{E}$ as the ``image'' of the previous $L$ items in the latent space and regard sequential patterns as local features of this ``image''.
This approach enables the use of convolution filters to search for sequential patterns. Figure~\ref{fig:hcase} shows two ``horizontal filters'' that capture two union-level sequential patterns. These filters, represented as $h\times d$ matrices, have the height $h=2$ and the full width equal to $d$. They pick up signals for sequential patterns by sliding over the rows of $\boldsymbol{E}$. For example, the first filter picks up the sequential pattern ``(Airport, Hotel) $\rightarrow$ Great Wall'' by having
larger values in the latent dimensions where Airport and Hotel have larger values.
Similarly, a ``vertical filter'' is a $L\times 1$ matrix and will slide over the columns of $\boldsymbol{E}$. More details are explained below. Unlike image recognition, the ``image'' $\boldsymbol{E}$  is not given because the embedding $\boldsymbol{Q}_i$ for all items $i$ must be learnt simultaneously with all filters.

\noindent{\textbf{Horizontal Convolutional Layer}}. This layer, shown in the upper part of the second component in Figure~\ref{fig:network}, has $n$ horizontal filters
$\boldsymbol{F}^k \in \mathbb{R}^{h \times d}$, $1\le k\le n$. $h \in \{1,\cdots, L\}$
is the height of a filter. For example, if $L=4$, one may choose to have $n=8$ filters, two for each $h$ in $\{1,2,3,4\}$. $\boldsymbol{F}^k$ will slide from top to bottom on $\boldsymbol{E}$ and interact with all horizontal dimensions of $\boldsymbol{E}$ of the items $i$, $1\leq i\leq L-h+1$. The result of the interaction
is the $i$-th convolution value given by
\begin{equation}\label{equ:hconv}
\boldsymbol{c}^k_i = \phi_{c} (\boldsymbol{E}_{i:i+h-1} \odot \boldsymbol{F}^k).
\end{equation}
where the symbol $\odot$ denotes the inner product operator and $\phi_c(\cdot)$ is the activation function for convolutional layers. This value is the inner product between $\boldsymbol{F}^k$ and the sub-matrix formed by the row $i$ to row $i-h+1$ of $\boldsymbol{E}$, denoted by $\boldsymbol{E}_{i:i+h-1}$.
The final convolution result of $\boldsymbol{F}^k$ is the vector
\begin{equation}
\boldsymbol{c}^k = \left[ \boldsymbol{c}^k_1~\boldsymbol{c}^k_2 \cdots \boldsymbol{c}^k_{L-h+1} \right].
\end{equation}
We then apply a max pooling operation to $\boldsymbol{c}^k$ to extract the maximum value from all values produced by this particular filter. The maximum value captures the most significant feature extracted by the filter. Therefore, for the $n$ filters in this layer, the output value $\boldsymbol{o} \in \mathbb{R}^n$ is
\begin{equation}
\boldsymbol{o} = \{max(\boldsymbol{c}^1), max(\boldsymbol{c}^2),\cdots,max(\boldsymbol{c}^n)\}.
\end{equation}

Horizontal filters interact with every successive $h$ items through their embeddings $\boldsymbol{E}$. Both the embeddings and the filters
are learnt to minimize an objective function that encodes the prediction error of target items (more in Section \ref{sec:training}). By sliding filters of various heights, a significant signal will be picked up regardless of location. Therefore, horizontal filters can be trained to capture \textbf{union-level patterns} with multiple union sizes.

\noindent{\textbf{Vertical Convolutional Layer.}}
This layer is shown in the lower part of the second component in Figure~\ref{fig:network}. We use tilde ($\sim$) for the symbols of this layer.
Suppose that there are $\tilde{n}$ vertical filters $\boldsymbol{\tilde{F}}^k \in \mathbb{R}^{L \times 1}$ , $1 \le k \le \tilde{n}$.
Each filter $\boldsymbol{\tilde{F}}^k$ interacts with the columns of $\boldsymbol{E}$
by sliding $d$ times from left to right on $\boldsymbol{E}$, yielding the vertical convolution result $\boldsymbol{\tilde{c}}^k$:
\begin{equation}
\boldsymbol{\tilde{c}}^k = \left[ \boldsymbol{\tilde{c}}^k_1~\boldsymbol{\tilde{c}}^k_2 \cdots \boldsymbol{\tilde{c}}^k_{d} \right].
\end{equation}
For the inner product interaction, it is easy to verify that this result is equal to the weighted sum over the $L$ rows of $\boldsymbol{E}$ with $\tilde{F}^k$ as the weights:
\begin{equation}\label{eqn:vertical}
\boldsymbol{\tilde{c}}^k=\sum_{l=1}^L \boldsymbol{\tilde{F}}^k_{l} \cdot \boldsymbol{E}_{l},
\end{equation}
where $\boldsymbol{E}_{l}$ is the $l$-th row of $\boldsymbol{E}$.
Therefore, with vertical filters we can learn to \emph{aggregate} the embeddings of
the $L$ previous items, similar to Fossil's~\cite{HeMcA16b} weighted sum to aggregate the $L$ previous items' latent representations. The difference is that each filter $\boldsymbol{\tilde{F}}^k$ is acting like a different aggregator.
Thus, similar to Fossil, these vertical filters are capturing \textbf{point-level sequential patterns} through weighted sums over previous items' latent representations.
While Fossil uses a single weighted sum for each user, we can use $\tilde{n}$ global vertical filters to produce $\tilde{n}$ weighted sums $\boldsymbol{\tilde{o}} \in \mathbb{R}^{d\tilde{n}}$ for all users:
\begin{equation}
\boldsymbol{\tilde{o}} = \left[ \boldsymbol{\tilde{c}}^1~\boldsymbol{\tilde{c}}^2 \cdots \boldsymbol{\tilde{c}}^{\tilde{n}} \right].
\end{equation}
Since their usage is aggregation, vertical filters have some differences from horizontal ones: (1) The size of each vertical filter is fixed to be $L \times 1$. This is because each column of $\boldsymbol{E}$ is latent for us, it is meaningless to interact with multiple successive columns at one time. (2) There is no need to apply max pooling operation over the vertical convolution results, as we want to keep the aggregation for every latent dimension.
Thus, the output of this layer is $\boldsymbol{\tilde{o}}$.
\subsection{Fully-connected Layers} 

We concatenate the outputs of the two convolutional layers and feed them into a fully-connected neural network layer to get more high-level and abstract features:
\begin{equation}
\boldsymbol{z} = \phi_{a} (\boldsymbol{W} \left[\begin{matrix}  \boldsymbol{o} \\
\boldsymbol{\tilde{o}} \end{matrix}\right] + \boldsymbol{b}),
\end{equation}
where $\boldsymbol{W} \in \mathbb{R}^{d \times (n+d\tilde{n})}$ is the weight matrix that projects the concatenation layer to a $d$-dimensional hidden layer, $\boldsymbol{b} \in \mathbb{R}^d$ is the corresponding bias term and $\phi_a(\cdot)$ is the activation function for fully-connected layer. $\boldsymbol{z} \in \mathbb{R}^d$ is what we called \emph{convolutional sequence embedding}, which encodes all kinds of sequential features of the $L$ previous items.

To capture user's general preferences, we also look-up the user embedding $\boldsymbol{P}_u$ and
concatenate the two $d$-dimensional vectors, $\boldsymbol{z}$ and $\boldsymbol{P}_u$, together and project them to an output layer with $|\mathcal{I}|$ nodes, written as
\begin{equation}
\boldsymbol{y}^{(u,t)} = \boldsymbol{W'} \left[\begin{matrix}  \boldsymbol{z} \\
\boldsymbol{P}_u \end{matrix}\right] + \boldsymbol{b'},
\end{equation}
where $\boldsymbol{b'} \in \mathbb{R}^{|\mathcal{I}|}$ and $\boldsymbol{W'} \in \mathbb{R}^{|\mathcal{I}| \times 2d}$ are the bias term and weight matrix for output layer, respectively. As explained in Section \ref{sec:training}, the value $\boldsymbol{y}^{(u,t)}_i$ in the output layer is associated with the probability of how likely user $u$ will interact with item $i$ at time step $t$.
$\boldsymbol{z}$ intends to capture short term sequential patterns, whereas
the user embedding $\boldsymbol{P}_u$ captures user's long-term general preferences.
Here we put the user embedding Pu in the last hidden layer for several reasons: (1) As we shall see in Section~\ref{sec:connect}, it can have the ability to generalize other models. (2) we can pre-train our model's parameters with other generalized models' parameters. As stated in~\cite{he2017neural}, such pre-training is critical to model performance


\subsection{Network Training}\label{sec:training} 
To train the network, we transform the values of the output layer, $\boldsymbol{y}^{(u,t)}$, to probabilities by:
\begin{equation}
p(\mathcal{S}^{u}_{t}~|~\mathcal{S}^{u}_{t-1},\mathcal{S}^{u}_{t-2},\cdots,\mathcal{S}^{u}_{t-L}) = \sigma(\boldsymbol{y}^{(u,t)}_{\mathcal{S}^u_t}),
\end{equation}
where $\sigma(x) = 1 /(1+e^{-x})$ is the \emph{sigmoid} function.
Let $\mathcal{C}^u = \{L+1,L+2,...,|\mathcal{S}^u|\}$ be the collection of time steps for which we would like to make predictions for user $u$. The likelihood of all sequences in the dataset is:
\begin{equation}
p(\mathcal{S}|\Theta) = \prod_u \prod_{t \in \mathcal{C}^u} \sigma(\boldsymbol{y}^{(u,t)}_{\mathcal{S}^u_t}) \prod_{j \ne \mathcal{S}^u_t}( 1 - \sigma(\boldsymbol{y}^{(u,t)}_{j})).
\end{equation}


To further capture \textbf{skip behaviors},
we could consider the next $T$ target items, $\mathcal{D}^u_t = \{\mathcal{S}^u_t, \mathcal{S}^u_{t+1},...,\mathcal{S}^u_{t+T}\}$,
at once by replacing the immediate next item  $\mathcal{S}^u_t$
in the above equation
 with $\mathcal{D}^u_t$. Taking the negative logarithm of likelihood, we get the objective function, also known as \emph{binary cross-entropy} loss:
\begin{equation}\label{eqn:objective}
 \ell = \sum_u \sum_{t \in \mathcal{C}^u} \sum_{i \in \mathcal{D}^u_t} -\mathrm{log}(\sigma(\boldsymbol{y}^{(u,t)}_{i})) + \sum_{j \ne i} -\mathrm{log}(1 - \sigma(\boldsymbol{y}^{(u,t)}_{j})).
\end{equation}
Following previous works~\cite{rendle2010factorizing,HeMcA16b,wu2016collaborative}, for each target item $i$, we randomly sample several (3 in our experiments) negative instances $j$ in the second term.

The \emph{model parameters} $\Theta=\{\boldsymbol{P},\boldsymbol{Q},\boldsymbol{F},\boldsymbol{\tilde{F}},\boldsymbol{W},\boldsymbol{W'},\boldsymbol{b},\boldsymbol{b'}\}$ are learned by minimizing the  objective function in Eqn \eqref{eqn:objective} on the training set, whereas the \emph{hyperparameters} (\emph{e.g.,} $d,n,\tilde{n},L,T$) are tuned on the validation set via grid search.
We adopt an variant of Stochastic Gradient Descent (SGD) called Adaptive Moment Estimation (Adam)~\cite{kingma2014adam} for faster convergence, with a batch size of 100. To control model complexity and avoid over-fitting, we use two kinds of regularization methods: the $L2$ Norm is applied for all model parameters and  \emph{Dropout}~\cite{Srivastava2014dropout} technique with $50\%$ drop ratio is used on fully-connected layers.
We implemented Caser with MatConvNet~\cite{vedaldi2015matconvnet}. The whole training time is proportional to the number of training instances. For example, it took around 1 hour for MovieLens data and 2 hours for Gowalla data, 2 hours for Foursquare and 1 hour for Tmall on a 4-cores i7 CPU and 32GB RAM machine. These times are comparable to Fossil's~\cite{HeMcA16b} running time and can be further reduced by using GPU.

\subsection{Recommendation}\label{sec:prediction}
After obtaining the trained neural network, to make recommendations for a user $u$ at time step $t$, we take $u$'s latent embedding $\boldsymbol{P}_u$ and extract his last $L$ items' embeddings given by Eqn \eqref{equ:embed} as the neural network input. We recommend the $N$ items that have the highest values in the output layer $\boldsymbol{y}$. The complexity for making recommendations to all users is $O(|\mathcal{U}||\mathcal{I}|d)$, where the complexity of convolution operations is ignored. 
Note that the number of target items $T$ is a hyperparameter used during the model training, whereas $N$ is the number of items recommended after the model is trained.

\subsection{Connection to Existing Models}\label{sec:connect} 
We show that Caser is a generalization of several previous models.

\emph{Caser vs. MF}.
By discarding all convolutional layers and all bias terms, our model becomes a vanilla LFM with user embeddings as user latent factors and its associated weights as item latent factors.
MF usually contains bias terms\footnote{Top-$N$ recommendation ranks the items for each user individually, which is invariant to user bias and global bias.}, which is $\boldsymbol{b'}$ in our model. After discarding all convolutional layers, the resulting model is the same as MF:
\begin{equation}
\boldsymbol{y}^u_i = \boldsymbol{W'}_{i} \left[\begin{matrix}  \boldsymbol{0} \\
\boldsymbol{P}_u \end{matrix}\right] + \boldsymbol{b'}_i.
\end{equation}

\emph{Caser vs. FPMC}.
FPMC fuses factorized first-order Markov chain with LFM and is optimized by Bayesian personalized ranking (BPR). Although Caser uses a different optimization criterion, \emph{i.e.,} the cross-entropy, it is able to generalize FPMC by copying the previous item's embedding to the hidden layer $\boldsymbol{z}$ and not using any bias terms: \begin{equation}
\boldsymbol{y}^{(u,t)}_i = \boldsymbol{W'}_i \left[\begin{matrix}  \boldsymbol{Q}_{\mathcal{S}^u_{t-1}} \\
\boldsymbol{P}_u \end{matrix}\right].
\end{equation}
As FPMC uses BPR as the criterion, our model is not exactly the same as FPMC. However, BPR is limited to have only 1 target and negative sample at each time step. Our cross-entropy loss does not have these limitations.

\emph{Caser vs. Fossil}.
By omitting the horizontal convolutional layer and using one vertical filter and copying the vertical convolution result $\boldsymbol{\tilde{c}}$ to the hidden layer $\boldsymbol{z}$, we get
\begin{equation}
\boldsymbol{y}^{(u,t)}_i = \boldsymbol{W'}_i \left[\begin{matrix}  \boldsymbol{\tilde{c}} \\
\boldsymbol{P}_u \end{matrix}\right]  + \boldsymbol{b'}_i.
\end{equation}
As discussed for Eqn \eqref{eqn:vertical}, this vertical filter serves as the weighted sum of the embeddings of the $L$ previous items, like in Fossil, though Fossil uses \emph{Similarity Model} instead of LFM and factorizes it in the same latent space as Markov model. Another difference is that Fossil uses one local weighting for each user while we use a number of global weighting through vertical filters.


\section{Experiments}\label{sec:exper}
\begin{table*}[t!]
\center
\caption{Statistics of the datasets}\label{tb:dataset}
\vspace{-0.4cm}
\setlength{\tabcolsep}{10pt}
\begin{tabular}{lccccc}
\toprule
\multirow{2}{4em}{\textbf{Datasets}} &
 \textbf{Sequential} & \multirow{2}{3em}{\textbf{\#users}} & \multirow{2}{3em}{\textbf{\#items}} & \textbf{avg. actions} & \multirow{2}{3.5em}{\textbf{Sparsity}}\\
& \textbf{Intensity} & & & \textbf{per user} &\\
\midrule
MovieLens & 0.3265 & 6.0k & 3.4k & 165.50 & 95.16\%\\
\midrule
Gowalla & 0.0748 & 13.1k & 14.0k & 40.74 & 99.71\%\\
\midrule
Foursquare & 0.0378 & 10.1k & 23.4k & 30.16 & 99.87\%\\
\midrule
Tmall & 0.0104 & 23.8k & 12.2k & 13.93 & 99.89\%\\
\bottomrule
\end{tabular}
\end{table*}

We compare \emph{Caser} with state-of-the-art methods. The source code of Caser and processed data sets are available online\footnote{https://github.com/graytowne/caser}.

\begin{table*}[t]
\caption{Performance comparison on the four data sets.}
\vspace{-0.4cm}
\setlength{\tabcolsep}{0.25cm}
\centering
\label{tb:perform}
\begin{tabular}{c|c|cccccccc}
\toprule
Dataset & Metric & POP & BPR & FMC & FPMC & Fossil & GRU4Rec & Caser & Improv.\\
\midrule
\multirow{6}{5em}{\centering \textit{MovieLens}}& Prec@1 & 0.1280 & 0.1478 & 0.1748 & 0.2022 & 0.2306 & \textbf{0.2515} & 0.2502 & -0.5\%\\
&Prec@5 & 0.1113 & 0.1288 & 0.1505 & 0.1659 & 0.2000 & 0.2146 & \textbf{0.2175} & 1.4\%\\
&Prec@10 & 0.1011 & 0.1193 & 0.1317 & 0.1460 & 0.1806 & 0.1916 & \textbf{0.1991} & 4.0\%\\
&Recall@1 & 0.0050 & 0.0070 & 0.0104 & 0.0118 & 0.0144 & \textbf{0.0153} & 0.0148 & -3.3\%\\
&Recall@5 & 0.0213 & 0.0312 & 0.0432 & 0.0468 & 0.0602 & 0.0629 & \textbf{0.0632} & 0.5\%\\
&Recall@10 & 0.0375 & 0.0560 & 0.0722 & 0.0777 & 0.1061 & 0.1093 & \textbf{0.1121} & 2.6\%\\
&MAP & 0.0687 & 0.0913 & 0.0949 & 0.1053 & 0.1354 & 0.1440 & \textbf{0.1507} & 4.7\%\\
\midrule
\multirow{6}{5em}{\centering \textit{Gowalla}} & Prec@1 & 0.0517 & 0.1640 & 0.1532 & 0.1555 & 0.1736 & 0.1050 & \textbf{0.1961} & 13.0\%\\
&Prec@5 & 0.0362 & 0.0983 & 0.0876 & 0.0936 & 0.1045 & 0.0721 & \textbf{0.1129} & 8.0\%\\
&Prec@10 & 0.0281 & 0.0726 & 0.0657 & 0.0698 & 0.0782 & 0.0571 & \textbf{0.0833} & 6.5\%\\
&Recall@1 & 0.0064 & 0.0250 & 0.0234 & 0.0256 & 0.0277 & 0.0155 & \textbf{0.0310} & 11.9\%\\
&Recall@5 & 0.0257 & 0.0743 & 0.0648 & 0.0722 & 0.0793 & 0.0529 & \textbf{0.0845} & 6.6\%\\
&Recall@10 & 0.0402 & 0.1077 & 0.0950 & 0.1059 & 0.1166 & 0.0826 & \textbf{0.1223} & 4.9\%\\
&MAP & 0.0229 & 0.0767 & 0.0711 & 0.0764 & 0.0848 & 0.0580 & \textbf{0.0928} & 9.4\%\\
\midrule
\multirow{6}{5em}{\centering \textit{Foursquare}} & Prec@1 & 0.1090 & 0.1233 & 0.0875 & 0.1081 & 0.1191 & 0.1018 & \textbf{0.1351} & 13.4\%\\
&Prec@5 & 0.0477 & 0.0543 & 0.0445 & 0.0555 & 0.0580 & 0.0475 & \textbf{0.0619} & 6.7\%\\
&Prec@10 & 0.0304 & 0.0348 & 0.0309 & 0.0385 & 0.0399 & 0.0331 & \textbf{0.0425} & 6.5\%\\
&Recall@1 & 0.0376 & 0.0445 & 0.0305 & 0.0440 & 0.0497 & 0.0369 & \textbf{0.0565} & 13.7\%\\
&Recall@5 & 0.0800 & 0.0888 & 0.0689 & 0.0959 & 0.0948 & 0.0770 & \textbf{0.1035} & 7.9\%\\
&Recall@10 & 0.0954 & 0.1061 & 0.0911 & 0.1200 & 0.1187 & 0.1011 & \textbf{0.1291} & 7.6\%\\
&MAP & 0.0636 & 0.0719 & 0.0571 & 0.0782 & 0.0823 & 0.0643 & \textbf{0.0909} & 10.4\%\\
\midrule
\multirow{6}{5em}{\centering \textit{Tmall}} & Prec@1 & 0.0010 & 0.0111 & 0.0197 & 0.0210 & 0.0280 & 0.0139 & \textbf{0.0312} & 11.4\%\\
&Prec@5 & 0.0009 & 0.0081 & 0.0114 & 0.0120 & 0.0149 & 0.0090 & \textbf{0.0179} & 20.1\%\\
&Prec@10 & 0.0007 & 0.0063 & 0.0084 & 0.0090 & 0.0104 & 0.0070 & \textbf{0.0132} & 26.9\%\\
&Recall@1 & 0.0004 & 0.0046 & 0.0079 & 0.0082 & 0.0117 & 0.0056 & \textbf{0.0130} & 11.1\%\\
&Recall@5 & 0.0019 & 0.0169 & 0.0226 & 0.0245 & 0.0306 & 0.0180 & \textbf{0.0366} & 19.6\%\\
&Recall@10 & 0.0026 & 0.0260 & 0.0333 & 0.0364 & 0.0425 & 0.0278 & \textbf{0.0534} & 25.6\%\\
&MAP & 0.0030 & 0.0145 & 0.0197 & 0.0212 & 0.0256 & 0.0164 & \textbf{0.0310} & 21.1\%\\
\bottomrule
\end{tabular}
\end{table*}

\subsection{Experimental Setup}
\noindent \textbf{Datasets}.
Sequential recommendation makes sense only when the data set contains sequential patterns. To identify such data sets, we applied sequential association rule mining to several public data sets and computed their sequential intensity defined by:
\begin{equation}\label{equ:SI}
\text{Sequential Intensity (\emph{SI})}=\frac{\text{\#rules}}{\text{\#users}}.
\end{equation}
The numerator is the total number of rules in the form of Eqn ~\eqref{equ:rules} found using a minimum threshold on support (\emph{i.e.,} 5) and confidence(\emph{i.e.,} 50\%) with Markov order $L$ range from 1 to 5. The denominator is the total number of users. We use $SI$ to estimate the intensity of sequential signals in a data set.

The four data sets with their $SI$ are described in Table~\ref{tb:dataset}. MovieLens\footnote{https://grouplens.org/datasets/movielens/1m/} is the widely used movie rating data. Gowalla\footnote{https://snap.stanford.edu/data/loc-gowalla.html} constructed by~\cite{cho2011friendship} and Foursquare obtained from  \cite{yuan2014graph} contain implicit feedback through user-venue check-ins. Tmall, the largest B2C platform in China, is a user-purchase data obtained from IJCAI 2015  competition\footnote{https://ijcai-15.org/index.php/repeat-buyers-prediction-competition}, which aims to forecast repeated buyers.
Following previous works \cite{HeMcA16b,rendle2009bpr,wu2016collaborative}, we converted all numeric ratings to implicit feedback of 1. We also removed cold-start users and items of having less than $n$ feedbacks, as dealing with cold-start recommendation is usually treated as a separate issue in the literature \cite{wu2016collaborative,he2017neural,HeMcA16b,rendle2010factorizing}. $n$ is 5,15,10,10 for MovieLens, Gowalla, Foursquare, and Tmall. The Amazon data previously used in \cite{HeMcA16b, HeKanMcA17} was not used due to its $SI$ (0.0026 for `Office Products' category, 0.0019 for `Clothing, Shoes, Jewelry' and 'Video Games' category), in other words, its
sequential signals are much weaker than the above data sets.


%

Following \cite{liu2009hybrid,zhao2016stellar,yuan2014graph},
we hold the first 70\% of actions in each user's sequence as the \emph{training set} and use the next 10\% of actions as the \emph{validation set} to search the optimal hyperparameter settings for all models. The remaining 20\% actions in each user's sequence are used as the \emph{test set} for evaluating a model's performance.

\noindent \textbf{Evaluation Metrics}.
As in ~\cite{pan2008one,rendle2010factorizing,wang2015collaborative,wu2016collaborative}, we evaluate a model by Precision@$N$, Recall@$N$, and Mean Average Precision (MAP). Given a list of top $N$ predicted items for a user, denoted $\hat{R}_{1:N}$, and the last 20\% of actions in her/his sequence (\emph{i.e.}, denoted  $R$ (\emph{i.e.}, the test set),
Precision@$N$ and Recall@$N$ are computed by
\begin{equation}\label{equ:pr_re}
\begin{aligned}
\text{Prec}@N=\frac{|R \bigcap \hat{R}_{1:N}|}{N},\\
\text{Recall}@N=\frac{|R \bigcap \hat{R}_{1:N}|}{|R|}.
\end{aligned}
\end{equation}
We report the average of these values of all users. $N\in \{1,5,10\}$.
The Average Precision (AP) is defined by
\begin{equation}\label{equ:map}
\text{AP}=\frac{\sum_{N=1}^{|\hat{R}|}\text{Prec}@N \times \text{rel}(N)}{|\hat{R}|},
\end{equation}
where $rel(N)=1$ if the $N$-th item in $\hat{R}$ is in $R$. The Mean Average Precision (MAP) is the average of AP for all users.



\begin{figure*}[!t]  
    \centering
    \includegraphics[scale=0.5]{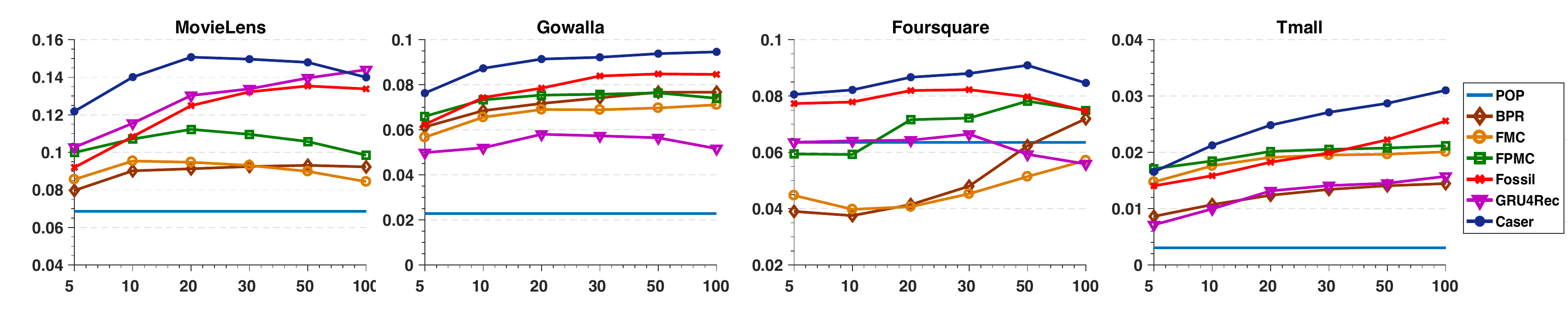}
    \vspace{-0.3cm}
    \caption{MAP (y-axis) vs. the number of latent dimensions $d$ (x-axis). }\label{fig:map_d}
\end{figure*}

\begin{figure*}[!t]  
    \vspace{-0.3cm}
    \centering
    \includegraphics[scale=0.5]{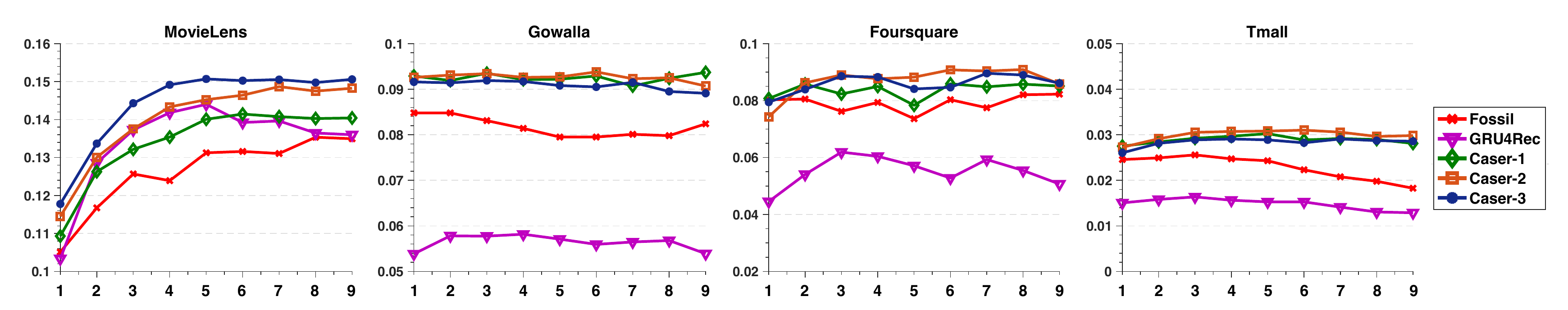}
    \vspace{-0.3cm}
    \caption{MAP (y-axis) vs. the Markov order $L$ (x-axis). Caser-1, Caser-2, and Caser-3 denote Caser with the number of targets $T$ set to $1,2,3$.}\label{fig:map_L}
\end{figure*}

\subsection{Performance Comparison}\label{sec:exper_compare}
We compare our method, Caser, proposed in Section \ref{sec:model} with the following baselines. 

\begin{itemize}
\item[$\bullet$] \textbf{POP}.
All items are ranked by their popularity in all users' sequences, and the popularity is determined by the number of interactions.

\item[$\bullet$] \textbf{BPR}.
Combined with Matrix Factorization model, Bayesian personalized ranking~\cite{rendle2009bpr} is the state-of-the-art method for non-sequential item recommendation on implicit feedback data.

\item[$\bullet$] \textbf{FMC and FPMC}.
As introduced in \cite{rendle2010factorizing}, FMC factorizes the first-order Markov transition matrix into two low-dimensional sub-matrices, and FPMC is a fusion of FMC and LFM. These are the state-of-the-art sequential recommendation methods. FPMC allows a basket of several items at each step. For our sequential recommendation problem, each basket has a single item.


\item[$\bullet$] \textbf{Fossil}.
Fossil~\cite{HeMcA16b} models high-order Markov chains and uses Similarity Model instead of LFM for modeling general user preferences.

\item[$\bullet$] \textbf{GRU4Rec}.
This is the session-based recommendation proposed by~\cite{hidasi2015session}. This model uses RNN to capture sequential dependencies and make predictions.

\end{itemize}

For each method, the grid search
is applied to find the optimal settings of hyperparameters using the validation set. These include latent dimensions $d$ from $\{5,10,20,30,50,100\}$, regularization hyperparameters, and the learning rate from $\{1, 10^{-1},..., 10^{-4}\}$. For Fossil, Caser and GRU4Rec,
the Markov order $L$ is from $\{1,\cdots,9\}$.
For Caser itself, the height $h$ of horizontal filters is from $\{1,\cdots, L\}$, the target number
$T$ is from $\{1,2,3\}$, the activation functions $\phi_a$ and $\phi_c$ are from $\{\emph{$identity$}, \emph{$sigmoid$}, \emph{tanh}, \emph{relu}\}$. For each height $h$, the number of horizontal filters is from $\{4, 8, 16, 32, 64\}$. The number of vertical filters is from $\{1, 2, 4, 8, 16\}$.
We report the result of each method under its optimal hyperparameter settings.


The best results of the six baselines and Caser are summarized in Table~\ref{tb:perform}.
The best performer on each row is highlighted in bold face.
The last column is the improvement of Caser relative to the best baseline, defined as $\frac{Caser-baseline}{baseline}$. Except for MovieLens,
Caser improved the best baseline on all $N$ tested by a large margin \emph{w.r.t.}
the three metrics. Among the baseline methods, the sequential recommenders (\emph{e.g.,} FPMC and Fossil) usually outperform non-sequential recommenders (\emph{i.e.,} BPR) on all data sets, suggesting the importance of considering sequential information. FPMC and Fossil outperform FMC on all data sets, suggesting the effectiveness of personalization. On MovieLens, GRU4Rec achieved a performance close to Caser's, but got a much worse performance on the other three data sets. In fact, MovieLens has more sequential signals than the other three data sets, thus, the RNN-based GRU4Rec could perform well on MovieLens but can easily get biased on training sets of the other three data sets despite the use of regularization and dropout as described in~\cite{hidasi2015session}. In addition, GRU4Rec's recommendation is session-based, instead of personalized, which enlarge the generalization error to some extent.


In the following studies, we examine the impact of the hyperparameters $d,L,T$ one at a time by holding the remaining hyperparameters at their optimal settings. We focus on MAP as it is an overall performance indicator and consistent with other metrics. 

\subsubsection{Influence of Latent Dimensionality $d$}
Figure~\ref{fig:map_d} shows MAP for various $d$ while keeping the other optimal hyperparameters unchanged. On the denser MovieLens, a larger $d$ does not always lead to a better model performance. A model achieves its best performance when $d$ is chosen properly and gets worse for a larger $d$ because of over-fitting. But for the other three sparser data sets, each model requires more latent dimensions to achieve their best results. For all data sets, Caser beats the strongest baseline performance by using a relatively small number of latent dimensions.

\subsubsection{Influence of Markov Order $L$ and Target Number $T$}
We vary $L$ to explore how much of Fossil, GRU4Rec and Caser can gain from high-order information while keeping other optimal hyperparameters unchanged.
Caser-1, Caser-2, and Caser-3 denote Caser with the target number $T$ at 1, 2, 3 to study the effect of skip behaviors. The results are shown in Figure~\ref{fig:map_L}. On the dense MovieLens, Caser best utilizes the extra information provided by a larger $L$ and Caser-3 performs the best, suggesting the benefits of skip steps. However, for the sparser data sets, all models do not consistently benefit from a larger $L$. This is reasonable, because for a sparse data set, a higher order Markov chain tends to introduce both extra information and more noises. In most cases, Caser-2 slightly outperforms the other models on these three data sets.

\begin{table}[h!]
\center
\caption{MAP vs. Caser Components}\label{tb:component}
\vspace{-0.4cm}
\setlength{\tabcolsep}{10pt}
\begin{tabular}{lcc}
\toprule
& \textbf{MovieLens} & {\textbf{Gowalla}}\\
\midrule
Caser-p & 0.0935 & 0.0777 \\
\midrule
Caser-h & 0.1304 & 0.0805 \\
\midrule
Caser-v & 0.1403 & 0.0841\\
\midrule
Caser-vh & 0.1448 & 0.0856 \\
\midrule
Caser-ph & 0.1372 & 0.0911 \\
\midrule
Caser-pv & 0.1494 & 0.0921 \\
\midrule
Caser-pvh & 0.1507 & 0.0928 \\
\bottomrule
\end{tabular}
\end{table}

\subsubsection{Analysis of Caser Components}
Finally, we evaluate the contribution of each of Caser's components,
the horizontal convolutional layer (\emph{i.e.}, $o$), the vertical convolutional layer (\emph{i.e.}, $\tilde{o}$), and personalization  (\emph{i.e.}, $P_u$),
to the overall performance while keeping all hyperparameters at their optimal settings. The result is shown in Table~\ref{tb:component} for MovieLens and Gowalla; the results of the other two data sets are similar. For $x \in \{p,h,v,vh,ph,pv,pvh\}$, Caser-$x$ denotes Caser with the components $x$ enabled. h denotes horizontal convolutional layer; v denotes vertical convolutional layer; p denotes personalization, which is similar to BPR and uses LFM only. Any missing component is represented by setting its corresponding $o$, $\tilde{o}$, and $P_u$ to zero. For example, vh denotes both vertical convolutional layer and horizontal convolutional layer by setting $P_u$ to all zeros, and pv denotes vertical convolutional layer and personalization by setting $o$ to all zeros.
Caser-p performs the worst whereas Caser-h, Caser-v, and Caser-vh  improve
the performance significantly, suggesting that treating top-$N$ sequential recommendation as the conventional top-$N$ recommendation will lose useful information, and that modeling both sequential patterns at the  union-level and point-level is useful for improving the prediction. 
For both data sets, the best performance is achieved by jointly using all parts of Caser, i.e., Caser-pvh.

\subsection{Network Visualization}
\begin{figure}[t]  
\centering
\includegraphics[scale=0.35]{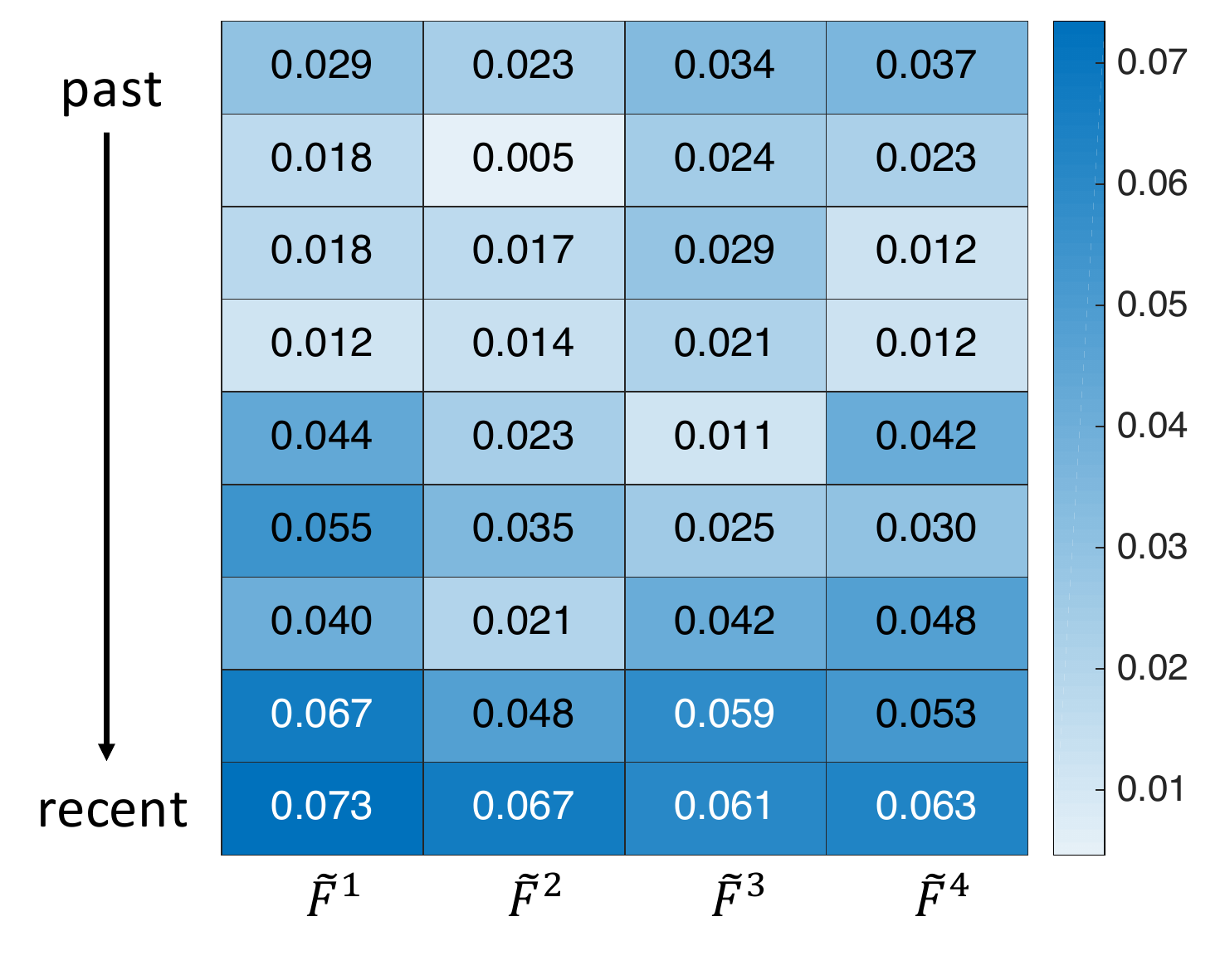}
\vspace{-0.5cm}
\caption{Visualization for four vertical convolutional filters of a trained model on MovieLens data when $L=9$.}
\label{fig:vis_vertical}
\end{figure}

We have a closer look at some trained networks and prediction. 
Figure~\ref{fig:vis_vertical} shows the values of four vertical convolutional filters after training Caser on MovieLens with $L=9$. In the micro perspective, the four filters are trained to be diverse, but in the macro perspective, they follow an ascending trend from past positions to recent positions. With each vertical filter serving as a way of
weighting the embeddings of previous actions (see the related discussion in Section \ref{sec:model}), this trend indicates that
Caser puts more emphasis on recent actions, demonstrating a major difference from the conventional top-$N$ recommendation.

\begin{figure}[t]  
\centering
\includegraphics[scale=0.11]{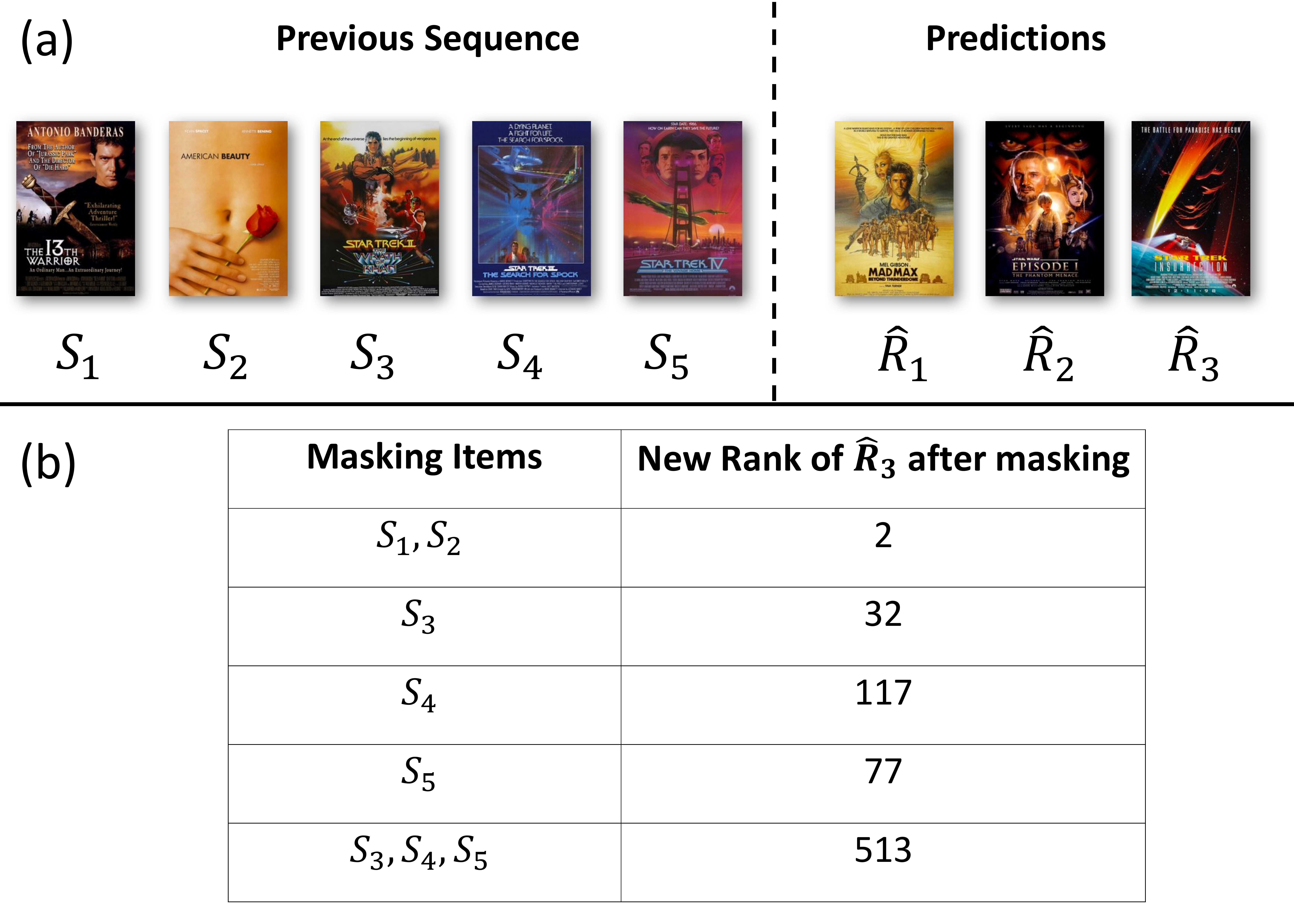}
\vspace{-0.3cm}
\caption{Horizontal convolutional filters's effectiveness of capturing union-level sequential patterns on MovieLens data.}
\label{fig:case}
\end{figure}

%

To see the effectiveness of horizontal filters, Figure~\ref{fig:case}(a) shows top $N=3$ ranked movies recommended by Caser, \emph{i.e.}, $\hat{R}_1$ (Mad Max), $\hat{R}_2$ (Star War), $\hat{R}_3$ (Star Trek) in that order, for a user with $L=5$ previous movies, \emph{i.e.}, $S_1$ (13th Warrior), $S_2$ (American Beauty), $S_3$ (Star Trek), $S_4$ (Star Trek III), and $S_5$ (Star Trek IV). $\hat{R}_3$ is the ground truth (\emph{i.e.}, the next movie in the user sequence). Note that $\hat{R}_1$ and $\hat{R}_2$ are quite similar to $\hat{R}_3$, \emph{i.e.}, all being action and science fiction movies, so are also recommended to the user.
Figure~\ref{fig:case}(b) shows the new rank of $\hat{R}_3$ after masking some of the $L$ previous movies by setting their item embeddings to zeros in the trained network. Masking $S_1$ and $S_2$ actually increases the rank of $\hat{R}_3$ to 2 (from 3); in fact, $S_1$ and $S_2$ are history or romance movies and act like noises for recommending $\hat{R}_3$. Masking each of $S_3$, $S_4$ and $S_5$ decreases the rank of $\hat{R}_3$ because these movies are in the same category as $\hat{R}_3$. The most decrease occurs after masking $S_3$, $S_4$ and $S_5$ all together. This study clearly indicates that our model correctly captures the dependence of $\hat{R}_3$ on the related $\{S_3, S_4, S_5\}$ as a union-level sequential feature for recommending $\hat{R}_3$.

\section{Conclusion}
Caser is a novel solution to top-$N$ sequential recommendation by modeling
recent actions as an ``image'' among time and latent dimensions
and learning sequential patterns using convolutional filters. This approach provides a unified and flexible
network structure for capturing many important features of sequential recommendation, \emph{i.e.}, point-level and union-level sequential patterns, skip behaviors, and long term user preferences. Our experiments and case studies on public real life data sets suggested that Caser outperforms the state-of-the-art methods for top-$N$ sequential recommendation.
%

\section*{Acknowledgement}
The work of the second author is partially supported by a Discovery Grant from Natural Sciences and Engineering Research Council of Canada. 

\bibliographystyle{ACM-Reference-Format}
\balance
{\small
\bibliography{abbrv_kdd15}
}

\end{document}